\documentclass[twocolumn,floatfix,superscriptaddress]{revtex4}
\usepackage{graphics}
\newcommand\etal{{\it et al.}\ }

\begin{document}

\title{Production of Antihydrogen at Reduced Magnetic Field for Anti-atom Trapping}
\author{G.B. Andresen}
\affiliation{Department of Physics and Astronomy, Aarhus University, DK-8000 Aarhus C, Denmark}
\author{W. Bertsche}
\affiliation{Department of Physics, Swansea University, Swansea SA2 8PP, United Kingdom}
\author{A. Boston}
\affiliation{Department of Physics, University of Liverpool, Liverpool L69 7ZE, United Kingdom}
\author{P.D. Bowe}
\affiliation{Department of Physics and Astronomy, Aarhus University, DK-8000 Aarhus C, Denmark}
\author{C.L. Cesar}
\affiliation{Instituto de F\'{i}sica, Universidade Federal do Rio de Janeiro, Rio de Janeiro 21941-972, Brazil}
\author{S. Chapman}
\affiliation{Department of Physics, University of California at Berkeley, Berkeley, CA 94720-7300, USA}
\author{M. Charlton}
\affiliation{Department of Physics, Swansea University, Swansea SA2 8PP, United Kingdom}
\author{M. Chartier}
\affiliation{Department of Physics, University of Liverpool, Liverpool L69 7ZE, United Kingdom}
\author{A. Deutsch}
\affiliation{Department of Physics, University of California at Berkeley, Berkeley, CA 94720-7300, USA}
\author{J. Fajans}
\affiliation{Department of Physics, University of California at Berkeley, Berkeley, CA 94720-7300, USA}
\author{M.C. Fujiwara}
\affiliation{TRIUMF, 4004 Wesbrook Mall, Vancouver BC, Canada V6T 2A3}
\author{R. Funakoshi}
\affiliation{Department of Physics, University of Tokyo, Tokyo 113-0033, Japan}
\author{D.R. Gill}
\affiliation{TRIUMF, 4004 Wesbrook Mall, Vancouver BC, Canada V6T 2A3}
\author{K. Gomberoff}
\affiliation{Department of Physics, University of California at Berkeley, Berkeley, CA 94720-7300, USA}
\author{J.S. Hangst}
\affiliation{Department of Physics and Astronomy, Aarhus University, DK-8000 Aarhus C, Denmark}
\author{R.S. Hayano}
\affiliation{Department of Physics, University of Tokyo, Tokyo 113-0033, Japan}
\author{R. Hydomako}
\affiliation{Department of Physics and Astronomy, University of Calgary, Calgary AB, Canada T2N 1N4}
\author{M.J. Jenkins}
\affiliation{Department of Physics, Swansea University, Swansea SA2 8PP, United Kingdom}
\author{L.V. J\o rgensen}
\affiliation{Department of Physics, Swansea University, Swansea SA2 8PP, United Kingdom}
\author{L. Kurchaninov}
\affiliation{TRIUMF, 4004 Wesbrook Mall, Vancouver BC, Canada V6T 2A3}
\author{N. Madsen}
\affiliation{Department of Physics, Swansea University, Swansea SA2 8PP, United Kingdom}
\author{P. Nolan}
\affiliation{Department of Physics, University of Liverpool, Liverpool L69 7ZE, United Kingdom}
\author{K. Olchanski}
\affiliation{TRIUMF, 4004 Wesbrook Mall, Vancouver BC, Canada V6T 2A3}
\author{A. Olin}
\affiliation{TRIUMF, 4004 Wesbrook Mall, Vancouver BC, Canada V6T 2A3}
\author{R.D. Page}
\affiliation{Department of Physics, University of Liverpool, Liverpool L69 7ZE, United Kingdom}
\author{A. Povilus}
\affiliation{Department of Physics, University of California at Berkeley, Berkeley, CA 94720-7300, USA}
\author{F. Robicheaux}
\affiliation{Department of Physics, Auburn University, Auburn, AL 36849-5311, USA}
\author{E. Sarid}
\affiliation{Department of Physics, NRCN-Nuclear Research Center Negev, Beer Sheva, IL-84190, Israel}
\author{D.M. Silveira}
\affiliation{Instituto de F\'{i}sica, Universidade Federal do Rio de Janeiro, Rio de Janeiro 21941-972, Brazil}
\author{J.W. Storey}
\affiliation{TRIUMF, 4004 Wesbrook Mall, Vancouver BC, Canada V6T 2A3}
\author{R.I. Thompson}
\affiliation{Department of Physics and Astronomy, University of Calgary, Calgary AB, Canada T2N 1N4}
\author{D.P. van der Werf}
\affiliation{Department of Physics, Swansea University, Swansea SA2 8PP, United Kingdom}
\author{J.S. Wurtele}
\affiliation{Department of Physics, University of California at Berkeley, Berkeley, CA 94720-7300, USA}
\author{Y. Yamazaki}
\affiliation{Atomic Physics Laboratory, RIKEN, Saitama 351-0198, Japan}
\collaboration{ALPHA Collaboration}
\noaffiliation

\date{\today}
\begin{abstract}  We have demonstrated production of antihydrogen in a 1$\,$T solenoidal magnetic field.  This field strength is significantly smaller than that used in the first generation experiments ATHENA (3$\,$T) and ATRAP (5$\,$T).   The motivation for using a smaller magnetic field is to facilitate trapping of antihydrogen atoms in a neutral atom trap surrounding the production region.  We report the results of measurements with the ALPHA (Antihydrogen Laser PHysics Apparatus) device, which can capture and cool antiprotons at 3$\,$T, and then mix the antiprotons with positrons at 1$\,$T.  We infer antihydrogen production from the time structure of antiproton annihilations during mixing, using mixing with heated positrons as the null experiment, as demonstrated in ATHENA.  Implications for antihydrogen trapping are discussed.
\end{abstract}

\maketitle

Cold antihydrogen atoms were first synthesized and detected in 2002 \cite{amor:02} by the ATHENA collaboration at the CERN Antiproton Decelerator (AD) \cite{maur:97}. The neutral antihydrogen atoms were not confined;  in fact, ATHENA detected the annihilation of the antiproton and positron in spatial and temporal coincidence to demonstrate antihydrogen production.  The ATRAP collaboration reported a similar result, using an indirect detection technique based on field ionization \cite{gabr:02}, shortly thereafter.  In both of the initial experiments, antihydrogen was produced by merging plasmas of antiprotons and positrons in liquid helium cooled Penning traps.  ATHENA observed peak antihydrogen production rates of up to about 400$\,$Hz \cite{amor:04}, immediately suggesting that an experiment to trap the neutral anti-atoms could be feasible.  Trapping of antihydrogen is probably necessary, if the long-term goal of performing precision spectroscopy of antihydrogen is to be realized.  Gravitational studies using antihydrogen will almost certainly require trapped anti-atoms.

We have constructed the first apparatus designed to produce and trap antihydrogen.  The ALPHA (Antihydrogen Laser PHysics Apparatus) device combines antihydrogen synthesis Penning traps with a superposed magnetic gradient trap for neutrals.  This device features a transverse octupole winding and a unique longitudinal magnetic field configuration involving multiple solenoidal windings \cite{bert:06}, designed to optimize antiproton capture, antihydrogen production rate, and antihydrogen trapping probability.  In this Letter, we demonstrate antihydrogen production at 1$\,$T in this multiple solenoid configuration.

\begin{figure}[hbt]
\centerline{\resizebox{\columnwidth}{!}{\includegraphics{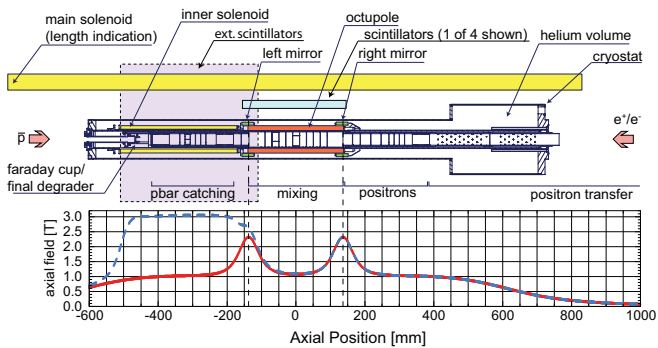}}}
\caption{Schematic diagram of the ALPHA apparatus.  The graph shows the on-axis longitudinal magnetic field due to the solenoids and mirror coils. The blue (red) curve is the field with (without) the inner solenoid energized.}
\label{fig1}
\end{figure}

Neutral atoms, or anti-atoms, can be trapped by exploiting the interaction of their magnetic dipole moments with an inhomogeneous magnetic field.  A potential well can be formed using a minimum-B configuration, as first described by Pritchard \cite{prit:83}.  The Ioffe-Pritchard configuration utilizes a cylindrical quadrupole for transverse confinement and solenoidal mirror coils for creating the longitudinal well.  The ALPHA apparatus, illustrated in Figure~\ref{fig1}, replaces the quadrupole with an octupole, in order to minimize perturbations that could lead to loss of the charged particle plasmas used to form antihydrogen.  Most laboratory Penning trap plasmas are stored in solenoidal fields having high uniformity and rotational symmetry, since the plasmas depend on this symmetry for their long-term stability \cite{onei:80}. The deleterious effects of a quadrupole field and the advantages of the octupole configuration are described elsewhere \cite{faja:04,faja:05,faja:06,gomb:07a}.  An earlier experiment in the ALPHA apparatus \cite{andr:07} showed that positrons and antiprotons can be stored in a strong octupole field for times comparable to those needed to produce antihydrogen in ATHENA.

The solenoidal field needed to confine charged antimatter particles represents a major challenge for the design of an effective antihydrogen trap.  The trap depth of a neutral trap is given by
 \begin{equation}U=\mu (B_{max}-B_{min}),
\label{1}
\end{equation}
where $\mu$ is the anti-atom's magnetic dipole moment and $B_{max}$ and $B_{min}$ are the maximum and minimum magnetic field strengths in the device.  In a combined Penning/neutral atom trap, the solenoidal field for the Penning trap is $B_{min}$.
Longitudinally, $B_{max}$ is given by
\begin{equation}B_{max}=B_{s}+B_{m},
\label{2}
\end{equation}
where $B_{s}$ is the solenoid field and $B_{m}$ is the peak field due to the mirror coil.  Transversely, we have
\begin{equation} B_{max}=\sqrt{{B_{s}}^{2}+{B_{w}}^{2}},
\label{3}
\end{equation}
where  $B_{w}$ is the transverse field strength of the multipole at the inner wall of the Penning trap.

The maximum trapping fields obtainable are fundamentally determined by the critical current in the superconductor used to generate the field.   The critical current is in turn larger for smaller external field strength.  Thus the solenoidal field should be as small as possible to maximize the trap depth.  Quantitatively, a trap depth of 1$\,$T provides about 0.7$\,$K of trapping potential for ground state antihydrogen.  (Note that the highly excited antihydrogen states observed in ATRAP and ATHENA may have significantly larger magnetic moments and thus be more trappable.  Cold rubidium atoms in highly excited Rydberg states have recently been trapped \cite{choi:05} in a superconducting Ioffe-Pritchard trap.)  Assuming that the maximum field strength in the superconductor is  4-5$\,$T, a background solenoidal field of 3 or 5$\,$T represents an undesirably large bias field for the trap.  The situation is exacerbated by the fact that the inner wall of the Penning trap is radially separated by a few mm from the innermost superconducting windings, due to the thickness of the magnet support structure and of the Penning trap itself. The loss of useful field strength in this distance is particularly significant for higher order multipole magnets.

In the absence of a neutral trap, a large solenoidal field is desirable for most aspects of the antihydrogen production cycle.  The antiprotons from the AD are slowed in a foil (final degrader in Figure~\ref{fig1}) from 5.3$\,$MeV to 5$\,$keV or less before trapping. The beam, which is partially focused by traversing the fringe field of the solenoid, has a transverse size of a few mm at the foil. Scattering in the foil adds divergence to the beam. The solenoidal field strength and the transverse size of the Penning trap electrodes (33.6$\,$mm diameter for the ALPHA catching trap) thus determine what fraction of the slowed particles can be transversely confined.  High magnetic field is also favored by considerations of cyclotron radiation cooling times for electrons and positrons, positron and antiproton plasma density (and thus antihydrogen production rate), and plasma storage lifetimes.

\begin{figure}[hbt]
\centerline{\resizebox{9cm}{!}{\includegraphics{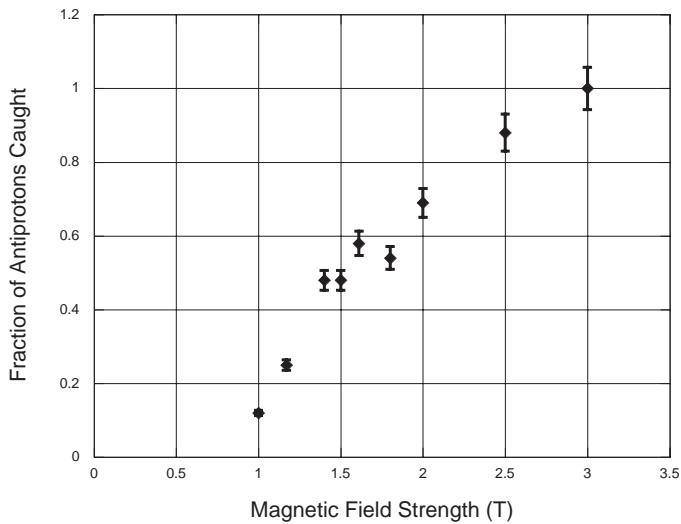}}}
\caption{Relative antiproton capture efficiency versus magnetic field strength. The measurements are relative to the result for 3$\,$T. The uncertainties reflect counting statistics only (1 standard deviation.)}
\label{fig2}
\end{figure}

In the following we concentrate on manipulations without the transverse octupole field energized.  A measurement of the relative antiproton capture efficiency versus solenoid field strength in ALPHA  is shown in Figure~\ref{fig2}.  For this measurement, the antiproton bunch from the AD, containing typically 2$\times$10$^{7}$ particles in 200$\,$ns, was slowed and trapped by pulsing the 5$\,$kV antiproton catching trap; see Figure~\ref{fig1}.  The "hot" antiprotons were then held for 500$\,$ms, before being released onto the final degrader (see Figure~\ref{fig1}), where they annihilate.  The annihilation products (charged pions) were counted using the external scintillation detectors (Figure~1). The magnetic field was provided by the ALPHA double solenoid system.  The main (external) solenoid was held at 1$\,$T, and the internal solenoid was varied from zero to 2$\,$T.  The 3$\,$T field is about a factor of eight more effective than a 1$\,$T field for capturing antiprotons, so the use of a single solenoid at low field for a combined apparatus seems ill advised. The ALPHA double solenoid is designed to catch antiprotons at 3$\,$T and to produce antihydrogen at 1$\,$T in the combined neutral/Penning trap.  In the following we demonstrate that the anticipated reductions in positron and antiproton density in the 1$\,$T field are not prohibitive for antihydrogen production.

For each mixing cycle with positrons to produce antihydrogen, three bunches of antiprotons from the AD  were captured, cooled through interactions with a previously loaded plasma of cold electrons, and then transferred (without electrons) to a potential well adjacent to the mixing region in the 1$\,$T field region; see Figure~\ref{fig1}.  The left mirror coil (adjacent to the inner solenoid) was energized to provide a smooth transition from the 3$\,$T region to the 1$\,$T region. This transfer was accomplished with typically less than 10\% loss in antiprotons. The antiprotons were then injected into the mixing region, which has the potential configuration of  a nested Penning trap \cite{gabr:88} (Figure 3a), containing positrons from the ALPHA positron accumulator \cite{jorg:05}.  Typical particle numbers were 7000 antiprotons injected into 30 million positrons. The entire trapping apparatus is cooled to 4 K by the cryostat for the inner superconducting magnets.

\begin{figure}[hbt]
\centerline{\resizebox{8cm}{!}{\includegraphics{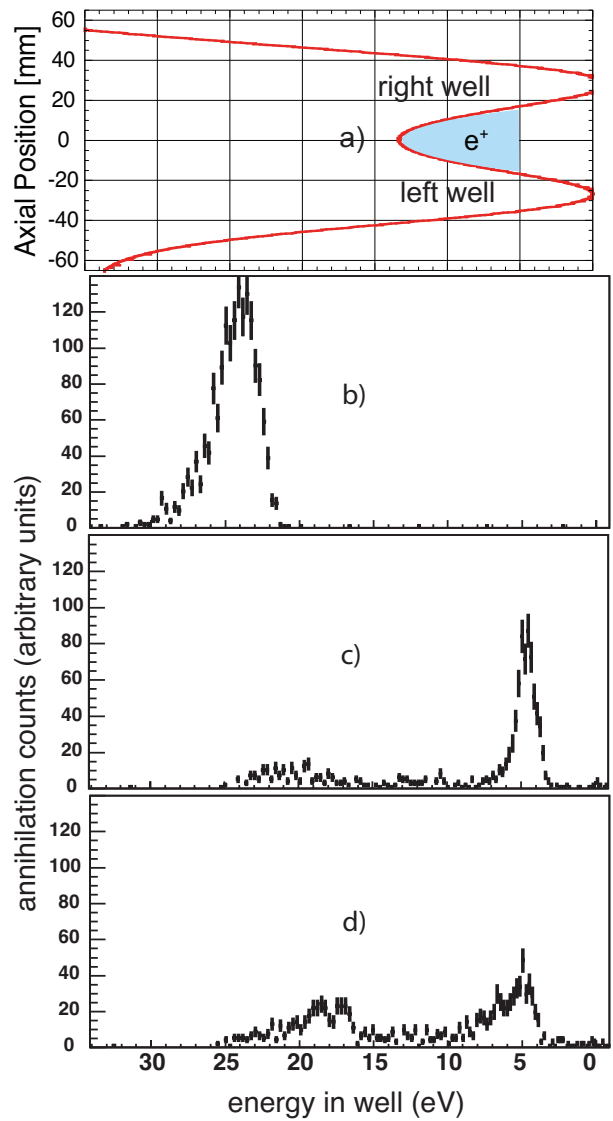}}}
\caption{a) The on-axis potential in the nested trap. The blue shaded region is the portion of the center well that is flattened by the positron space charge potential. b-d) Antiproton energy distributions in the nested trap potential measured by ramping down the left potential wall. The relative number of released antiprotons is plotted versus energy for b) antiprotons only, c) normal mixing with cold positrons, and d) mixing with heated positrons.  In all three cases, the antiprotons were released in 200$\,$ms after 50$\,$s of storage in the mixing trap. The horizontal axis scale is common to all four figures. The uncertainties reflect counting statistics only (1 standard deviation.)}
\label{fig3}
\end{figure}

The antiprotons, which are injected into the positron plasma with a relative energy of about 12 eV, slow by Coulomb interaction with the positrons, as previously observed in ATHENA \cite{amor:04b} and ATRAP \cite{gabr:01}.  The result of slowing can be observed by ramping down the trapping potential to determine at what energy the antiprotons are released. Figure~\ref{fig3} demonstrates positron cooling of antiprotons at 1$\,$T in ALPHA.  With no positrons, the antiprotons remain at the injection energy (Figure 3b). With positrons present, the antiprotons cool to an energy approximately corresponding to the potential at which the positron plasma is held (Figure 3c).  In ATHENA, cooling to this level was correlated with the onset of antihydrogen production \cite{amor:04b}, as measured by the rise in event rate in an antiproton annihilation detector.  The neutral antihydrogen escapes the Penning trap and annihilates on the electrode walls.

For the following measurements, the apparatus was equipped with four scintillation detectors read out by avalanche photodiodes. The detectors were placed inside the outer solenoid and adjacent to the mixing trap (Figure~\ref{fig1}). An event was registered if two or more of the detectors fired in coincidence (100$\,$ns window). The solid angle subtended by the detectors was about 35\% of 4$\pi$.

\begin{figure}[hbt]
\centerline{\resizebox{9cm}{!}{\includegraphics{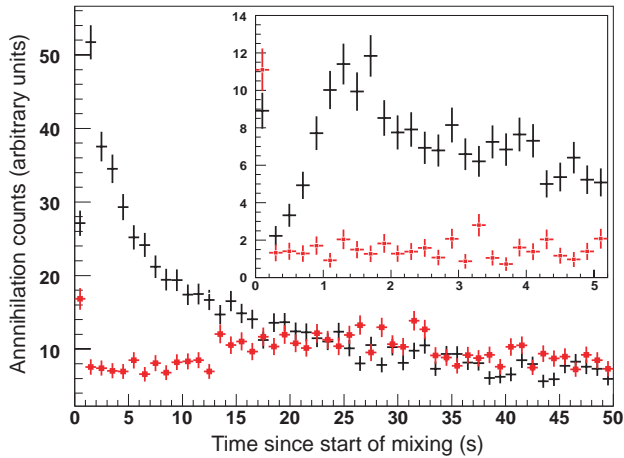}}}
\caption{Scintillation events as a function of time after the start of mixing, for normal mixing (black) and mixing with heated positrons (red). The time bins are 1$\,$s long. The data are for 10 mixing cycles, normalized to one cycle. The inset is a plot of the first 5$\,$s of the same data, re-binned into 200$\,$ms bins to illustrate the rise time of the antihydrogen production. The uncertainties reflect counting statistics only (1 standard deviation.) }
\label{fig4}
\end{figure}

Figure~\ref{fig4} illustrates the time development of the annihilation event rate after the start of mixing.  Two cases are shown; "normal" mixing and mixing in which the positrons are heated to suppress antihydrogen formation \cite{amor:02}. The heating is achieved by exciting the axial dipole mode of the positron plasma, again following established practice from ATHENA \cite{amor:03}. In normal mixing we observe the initial rise in event rate, as seen in the ATHENA apparatus, but with a considerably slower rise time - about 1$\,$s here as opposed to a few tens of ms.  This longer cooling time is probably due to the lower positron plasma density in the 1$\,$T field, although we have not measured the density directly.  The positron number here is also lower, by a factor of 2 to 3, than in \cite{amor:04b}.

The ATHENA experiment used position sensitive detection of antiproton and positron annihilation products to obtain the very first evidence for antihydrogen production at the AD.  In subsequent experiments, experience with the device demonstrated that it was not necessary to rely on the position-sensitive detection to distinguish antihydrogen production from antiproton loss { \cite{amor:04, amor:04c, amor:06}. The trigger rate signal from the annihilation detector exhibits a time structure that, in concert with evidence of antiproton cooling, can be interpreted as a signature for antihydrogen production.   Mixing with heated positrons leads to inefficient slowing and cooling of the antiprotons and inhibits antihydrogen production, and thus can serve as the null experiment.  In ALPHA, as in ATHENA, no evidence for significant antihydrogen production or significant antiproton loss is seen with heated positrons, although both species of particle are present and spatially overlapping during the cycle.  (The events in the very first time bin, for both cases, include "hot" antiproton losses caused by the rapid potential manipulations used to inject the particles into the nested trap.) We thus interpret the annihilation signal for cold mixing as being due to a time-varying antihydrogen production superimposed on a largely flat background due to cosmic rays and slow and small antiproton losses.  (There may be a small admixture of antihydrogen production even with heated positrons, at times greater than about 12$\,$s, but we have not yet investigated this in detail.)

Based on a knowledge of the number of antiprotons typically injected into the mixing trap, and the number remaining when the trap is dumped at the end of the cycle, we estimate that up to 15\% of the antiprotons could have produced antihydrogen.  This number is consistent with the total number of events observed, given the estimated scintillator detector efficiency, and it is comparable to that observed under typical conditions in ATHENA \cite{amor:04}.

The observation of antihydrogen produced in a 1$\,$T field is a significant development for the future of antihydrogen trapping experiments.  For example, the design of the ALPHA apparatus is for a maximum of 1.91$\,$T of transverse field from the octupole in a 1$\,$T solenoid, corresponding to a well depth of  1.16$\,$T. The well depth for a 3$\,$T solenoidal field and the same superconducting magnet construction technique \cite{bert:06} would be less than 0.5$\,$T, when the reduction in critical current is taken into account.  The relative ease with which antihydrogen was produced here suggests that attempts at even lower solenoid fields may succeed, leading to even larger neutral well depths. For possible work at lower field, the ALPHA device features the capability of applying rotating wall electric fields \cite{huan:97,andr:07a} to compress the antiproton and positron cloud radii before mixing, if necessary.

In summary, we have shown that antiprotons can be captured at high magnetic field, transferred to lower field without significant loss and then used to make antihydrogen, without further manipulation of the antiproton cloud.  This method is superior to performing the whole process at the lower field, and allows for a significantly higher neutral well depth for future attempts at antihydrogen trapping.

This work was supported by CNPq, FINEP (Brazil), ISF (Israel), MEXT (Japan), FNU (Denmark), NSERC, NRC (Canada), DOE, NSF (USA) and EPSRC (UK).



\providecommand{\newblock}{}

\end{document}